\documentclass{emulateapj}
\slugcomment{ApJ, in press}
\shorttitle{The X-ray Counterpart of PSR~J0726$-$2612}
\shortauthors{Speagle et al.}

\newcommand{\psr}{\object[PSR J0726-2612]{PSR J0726$-$2612}}
\newcommand{\chandra}{\textit{Chandra}}
\newcommand{\PSR}{\object[PSR J0726-2612]{PSR J0726}}
\newcommand{\expnt}[2]{\ensuremath{#1 \times 10^{#2}}}   
\newcommand{\rxj}{\object[RX J0420.0-5022]{RX J0420.0$-$5022}}

\newcommand{\rxjk}{\object[RX J0720.4-3125]{RX J0720.4$-$3125}}
\newcommand{\rbs}{\object{RX J1308.6+2127}}

\newcommand{\psrb}{\object[PSR J1718-3817]{PSR J1718$-$3817}}
\newcommand{\rxs}{\object[1RXS J072559.8-261229]{1RXS J072559.8$-$261229}}

\begin{document}

\title{The X-ray Counterpart of the High-$B$ Pulsar PSR J0726$-$2612}

\author{J.~S.~Speagle\altaffilmark{1,2}, D.~L.~Kaplan\altaffilmark{1,3}, 
  and M.~H.~van Kerkwijk\altaffilmark{4}}

\altaffiltext{1}{Physics Dept., U. of Wisconsin - Milwaukee, Milwaukee
  WI 53211, USA; kaplan@uwm.edu}
\altaffiltext{2}{Current address: Harvard College, Cambridge, MA
  02138, USA; joshuaspeagle@college.harvard.edu}
\altaffiltext{3}{Corresponding author.}
\altaffiltext{4}{Department of Astronomy and Astrophysics, University
  of Toronto, 60 St.\ George Street, Toronto, ON M5S 3H8, Canada;
mhvk@astro.utoronto.ca}

\begin{abstract}
  Middle-aged, cooling neutron stars are observed both as relatively
  rapidly spinning radio pulsars and as more slowly spinning, strongly
  magnetized isolated neutron stars (INSs), which stand out by their
  thermal X-ray spectra.  The difference between the two classes may
  be that the INSs initially had much stronger magnetic fields, which
  decayed.  To test this, we used the {\em Chandra X-ray Observatory}
  to observe \rxs, a possible X-ray counterpart to \psr, which, with
  its $3.44\,$s period and $\expnt{3}{13}\,$G inferred magnetic field
  strength, is the nearest and least extincted among the possible
  slowly-spinning, strong-field INS progenitors (it likely is in the
  Gould Belt, at $\sim\!1{\rm\,kpc}$).  We confirm the identification
  and find that the pulsar has a spectrum consistent with being purely
  thermal, with blackbody temperature $kT=87\pm5\,$eV and radius
  $R=5.7^{+2.6}_{-1.3}{\rm\,km}$ at a distance of 1\,kpc.  We detect
  sinusoidal pulsations at twice the radio period with a
  semi-amplitude of $27\pm5\%$.  The properties of \psr\ strongly
  resemble those of the INSs, except for its much shorter
  characteristic age of 200\,kyr (instead of several Myr).  We
  conclude that \psr\ is indeed an example of a young INS, one that
  started with a magnetic field strength on the low end of those
  inferred for the INSs, and that, therefore, decayed by a relatively
  small amount.  Our results suggest that the long-period,
  strong-field pulsars and the INSs are members of the same class, and
  open up new opportunities to understand the puzzling X-ray and
  optical emission of the INSs through radio observations of \psr.
\end{abstract}


\section{Introduction}
The \textit{ROSAT} All-Sky Survey (RASS; \citealt{rbs}) showed that
our census of cooling, nearby neutron stars was incomplete: it
contained not just the known cooling pulsars such as \object{PSR
  B0656+14} and \object{Geminga}, but also seven ``isolated neutron
stars'' (INSs).  These are nearby ($\lesssim\!1{\rm\,kpc}$), young
($\lesssim\!1{\rm\,Myr}$), cooling neutron stars with thermal X-ray
spectra, long periods ($>\!3{\rm\,s}$), faint optical counterparts,
and no detected radio emission (for reviews, see \citealt{haberl07,
  kaplan08}).  The INSs are interesting both because of their
abundance and because of the promise of inferring neutron-star
parameters from their thermal emission.  Unfortunately, despite large
investments of time with \chandra\ and \emph{XMM-Newton}, the nature
of the emission remains puzzling, and we still understand neither the
composition nor state (gaseous, condensed) of the surface.

A clearer picture has emerged for the origin and abundance of the
INSs: most likely, they were born with very strong magnetic fields, of
$\ga\!10^{13.5}{\rm\,G}$, which decayed.  Empirical evidence for this
comes from our X-ray timing efforts \citep{kvk05, kvk05b, kvk09,
  kvk09b, kvk11, vkk08}, which showed that the current field strengths
of the INSs are remarkably similar, in the range 1.0 to
$3.5\times10^{13}$\,G, and that their characteristic ages of several
Myr are substantially in excess of true ages of $\sim\!0.5$\,Myr
inferred from cooling and kinematics (\citealt{wal01,msh+05};
\citealt*{kvka07}; \citealt{vkk08, mph+09, tnhm10,tenh11}).  The long
periods and characteristic ages follow naturally if the INSs initially
had much stronger fields and thus faster spindown, and the similar
current field strengths can be understood if fields stop decaying at a
common value.  Indeed, this was predicted theoretically by
\citet*{pmg09}: for initially weak magnetic fields, field decay leads
to only a factor $\sim\!2$ change that is essentially unnoticeable
\citep{ppm+10}, while for fields above $10^{13}$\,G field decay
becomes increasingly important, with predicted final fields that are
always a ${\rm few}\times10^{13}$\,G, independent of the initial
values.  The field-decay induced heating may also help explain the
observed preponderance of INSs compared to ``normal'' middle-aged
pulsars \citep{kvk09, fgk06,kaspi10}.

If the above is correct, the progenitors of the INSs would have been
neutron stars with fields ranging from $\sim\!5$ to
$50\times10^{13}$\,G. The most magnetized would correspond to
magnetars, but many would not be so energetic. Still, as their fields
decay on $\sim\!100$\,kyr timescales, they should be brighter than
expected from simple cooling, and there should be a population of
relatively long-period, strong-field sources that are anomalously
hot. Verifying this might not only confirm the hypothesis for the INS
population, but also yield clues to their emission: with the higher
temperatures and stronger magnetic fields, one might expect the
surface to have a different state and the spectra to show different
spectral features that could be contrasted to the INSs.

\begin{figure}
\plotone{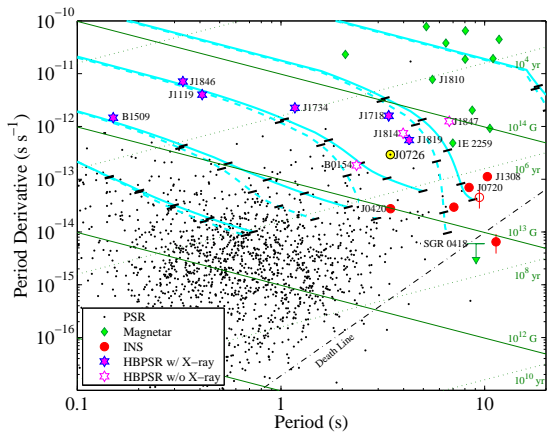}
\caption[]{$P$-$\dot P$ diagram, with known pulsars shown by dots, and
  strong-field populations highlighted (isolated neutron stars: red
  circles; magnetars: green diamonds; strong-field pulsars with X-ray
  observations: magenta stars [filled: detected; open: upper limit
    only]).  Sources referred to in the text are labeled, as are
  lines of constant magnetic field, lines of constant characteristic
  age, and an approximate death line.  Overlaid on the diagram are the
  predicted spin-down tracks from the field decay model of
  \citet{pmg09} (solid lines: dipole only; dashed lines: including a
  toroidal component with a strength of 50 times that of the dipole).
  The initial dipole fields are $10^{13,13.5,14,14.5,15}$\,G at the
  pole, and there are cross-ticks at $(0.01,0.03,0.1,0.3,1)\,$Myr.  We
  assumed initial periods of 0.1\,s, and divided the field at the pole
  by two to get the field at the equator for comparison with the
  spin-down field estimates.  For low field strengths, field decay is
  not important, and hence the tracks essentially follow
  constant-field lines (shown by the diagonal thin solid lines), and
  at any moment the characteristic age (diagonal thin dotted lines) is
  a reasonable estimate for the true age.  For strong fields, however,
  field decay is important, and the field and characteristic age
  inferred at any moment are not representative for the initial field
  and true age.}
\label{fig:ppdot}
\end{figure}

Checking the $P-\dot P$ diagram for possible INS progenitors
(Fig.~\ref{fig:ppdot}), one finds four radio pulsars, one rotating
radio transient (RRAT), and one relatively low-field anomalous X-ray
pulsar (AXP); all have periods longer than 3\,s and fields in excess
of $3\times10^{13}\,$G.  Of course, the AXP (\object{1E 2259+586}),
like other magnetars, is thought to be powered by magnetic field
decay; it certainly is anomalously hot, with $kT\simeq0.4\,$keV \citep{zkd+08}.  Of
the others, the RRAT and one of the pulsars have X-ray counterparts as
well, with thermal spectra and inferred temperatures of
$140-190\,$eV (\psrb, \citealt{km05,zkm+11}; \object[PSR
J1819-1458]{PSR~J1819$-$1458}, \citealt{mrg+07}).  These temperatures
are, as the authors mention, well above expectations from simple
cooling (and also well above the limit of $\sim\!70\,$eV inferred for
\object{PSR B0154+61} \citep{gkl+04}, a young pulsar with a somewhat
weaker inferred field; see Fig.~\ref{fig:ppdot}).  The best-studied
source, \object[PSR J1819-1458]{PSR~J1819$-$1458}, also has a clear
absorption feature at 1\,keV, well above the energies at which
absorption is seen in the INSs.  Among the three remaining sources,
\object[PSR J1814-1744]{PSR~J1814$-$1744} and \object[PSR
J1847-0130]{PSR~J1847$-$0130} have unpublished, reasonably long
\textit{XMM} observations.  We checked these and found no
counterparts, but as both pulsars are distant and extincted, this is
not unexpected.  Indeed all of these objects are considerably further
than the INSs, and while part of this can be explained by their small
ages leading to lower space densities, part comes from the
narrowly-beamed/intermittent radio emission that is used to detect
them (compared to omni-directional soft X-rays).

The last source, \psr\ (hereafter \PSR) is the subject of this
paper. This 3.44\,s pulsar was discovered in the course of the Parkes
High-Latitude survey \citep{burg+06} but surprisingly has not seen any
X-ray follow-up, despite its inferred magnetic field of
$B=3\times10^{13}\,$G and characteristic age of only $\tau\equiv
P/2\dot P=200\,$kyr.  Yet, it is arguably the most interesting, since
it should be the least extincted --- both because it has the lowest
dispersion measure (DM) of all (70 vs.\ 200 to $800{\rm\,cm^{-3}\,pc}$
for the other four pulsars) and because it is at what is, for a young
pulsar, high Galactic latitude ($|b^{II}|=4.7^\circ$ vs.\
$<\!0.22^\circ$ for the others).  Furthermore, it may well be the
closest.  From its dispersion measure, combined with a model of the
Galactic electron distribution \citep{cl02}, one infers a rough
distance of~3\,kpc.  But at that distance, its height above the
Galactic plane is 250\,pc, which is somewhat improbable.  A more
likely alternative is that it is at $\sim\!1\,$kpc, situated (and
born) in the \object{Gould Belt} \citep{ptp+05}, which this line of
sight passes through (and which may influence the dispersion measure).
Below, we will scale our distances to $d_{\rm kpc}=d/1{\rm\,kpc}$.

Overall, \PSR\ seems a prime candidate for comparison with the INSs,
being relatively close and unabsorbed, and apparently intermediate
between the stronger-field pulsars that look most like INS progenitors
and the weak-field normal pulsars that form the bulk of the
population.  Intriguingly, we found a possible counterpart to \PSR\ in
the RASS: \rxs, with $0.027\pm0.011{\rm\,ct\,s^{-1}}$.  The nominal
separation of $112''$ from the pulsar is relatively large, but in the
RASS image, the source seemed somewhat extended, encompassing the
pulsar.  Here, we present a {\em Chandra X-ray Observatory} of \rxs\
that confirms the identification with \PSR, and allows a first
comparison between \PSR\ and the INSs, opening the way for detailed
followup.

\section{Observations \& Analysis}
We observed \PSR\ with the Advanced CCD Imaging Spectrometer (ACIS;
\citealt{gbf+03}) aboard the \emph{Chandra X-ray Observatory}
(\emph{CXO}; \citealt{wtvso00}) on 2011 June 15 for 17.9\,ks (\dataset[ADS/Sa.CXO#obs/12558
]{ObsID 12558}).  For
best sensitivity at low energies, we used the back-illuminated S3 CCD,
selecting the 1/8-subarray read-out mode (with 0.4\,s sampling) to
resolve the 3-s pulse period.  We processed the level-1 event lists
following standard procedures (using \texttt{chandra\_repro} from CIAO
v4.3 and CALDB v4.4.5).  The data show one clear source, with a J2000
position (from \texttt{celldetect}) $\alpha=07^{\rm h}26^{\rm
  m}08\fs14$, $\delta=-26\degr12\arcmin38\farcs7$ with an uncertainty
dominated by the boresight error of $\sim\!0\farcs6$ (90\%
confidence).  This is consistent with the radio position of \PSR\ in
the ATNF pulsar catalog \citep{mhth05}): $\alpha=07^{\rm h}26^{\rm
  m}08\fs12\pm0\fs04$, $\delta=-26\degr12\arcmin38\farcs1\pm0\farcs8$.

For our further analysis, we selected 1179 source events from a
circular region with a radius of 5 pixels ($2\farcs46$), encompassing
$>95\%$ of the source photons, and 326 background events from the rest
of the detector area (a factor of 1200 larger in area), with energies
between 220\,eV (the recommended lower-bound for data from the ACIS;
any lower and the calibration becomes unreliable and soft flares
become increasingly dominant) and 1.1\,keV (where the effects of
higher-energy flares are minimized; the source is barely detected
above 1.1\,keV).  We corrected the event times to the Solar System
barycenter using \texttt{axbary}, assuming the radio position.

\begin{figure}
\plotone{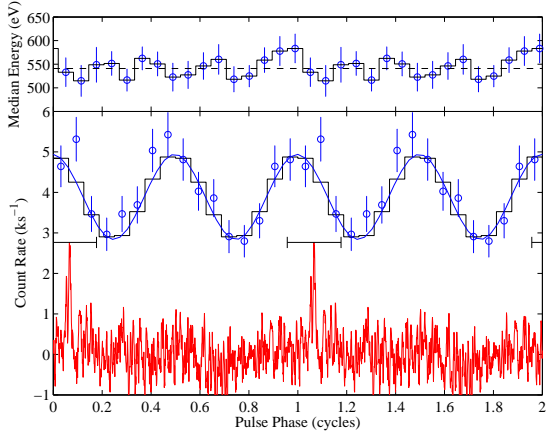}
\caption{Lower panel: X-ray (blue) and radio (red) pulse profiles of
  PSR J0726$-$2612.  The X-ray data are from our new
  \chandra\ observations, while the radio data are from the ATNF
  pulsar data archive with an arbitrary flux offset and scaling (these
  are from a 152-s observation centered at 1374\,MHz with 288\,MHz
  bandwidth, taken on 2005~March~26).  Both pulses are repeated twice
  for clarity.  The uncertainty on the X-ray TOA is 0.013\,cycles, but
  in comparing that to the radio TOA we must include the uncertainties
  on the ephemeris which contribute substantially.  The apparent shift
  between the radio and X-ray pulses is $0.07\pm0.11$\,cycles, as
  shown by the black error bars.  Upper panel: the median energy in
  each phase bin.  }
\label{fig:pulse}
\end{figure}

\subsection{Timing Analysis}
To look for X-ray pulsations, we first determined the frequency that
maximized the power in a $Z_1^{2}$ periodogram (Rayleigh statistic;
\citealt{bbb+83}) at different multiples of the radio frequency. We
found no obvious peak at the expected frequency of the pulsar
($Z_1^2=0.7$), but a strong one ($Z_1^2=21.7$) at roughly double the
frequency, $0.290493\pm0.000002$\,Hz (where the uncertainty was
calculated following \citealt{ransom01}).  We found no evidence for
significant power in higher harmonics ($Z_1^{2}=1.2$ and 0.70 at the
third and fourth harmonic, respectively).  We folded our data on half
our best-fit frequency to construct a binned light-curve (with 16
phase bins; Figure~\ref{fig:pulse}).  As expected from the lack of
other harmonics, a sinusoid provided a good fit ($\chi^{2}=11.7$ for
13 degrees of freedom [dof]), with a pulsed fraction (semi-amplitude)
of $27\%\pm5\%$.  The implied TOA, defined as the maximum of the
sinusoid closest to the middle of the observation, is MJD
$55727.6883125\pm0.0000006$.  The ephemeris in the ATNF catalog (based
on data from 2005--2006) predicts a frequency of
$0.29049679119\pm9.6\times10^{-10}$\,Hz at the epoch of our observation,
which differs by $\sim\!2\,\sigma$ from our measurement.  Further
radio observations would help determine whether this is due to chance
or due to something intrinsic (a glitch or timing noise).

In order to compare the phase of the radio and X-ray pulses, we
retrieved an archived Parkes observation of \PSR\ from 2005~March~26.
The pulse profile, shown in Fig.~\ref{fig:pulse}, was substantially
noisier than that shown in \citet{burg+06}, but still sufficed to
determine a TOA.  From the highest point in the profile, we find MJD
$53455.29291874\pm0.00000004$, with the uncertainty being $\pm 1$
phase bins.  With this TOA and the ATNF ephemeris, we infer a phase
difference of ${\rm TOA}_{\rm radio}-{\rm TOA_{\rm
    X}}=0.07\pm0.11$\,cycles, where the uncertainty is dominated by
uncertainties in the ephemeris (see Figure~\ref{fig:pulse}).

\subsection{Spectral Analysis}

Spectral fitting was done in \texttt{sherpa} \citep{sherpa}.  For the
fits, we only included events between 0.32--1.1\,keV, as the response
below 320\,eV is not well understood.  This leaves 1003 source and 267
background counts.  We follow recommendations in modeling the
low-energy background with a power-law instead of subtracting it
(although given that the expected background in the source area is
less than 1 count, it does not influence the results).  We binned the
source counts such that each bin contained at least 25 counts and was
at least 29\,eV wide, so that $\chi^2$ statistics are a good
approximation, and we do not oversample the instrumental resolution of
$\sim\!100\,$eV.


We tried three main source models: a thermal blackbody, a non-thermal
power law, and a simple neutron-star atmosphere model (NSA;
\citealt*{psz91,zps96}), all modified by interstellar absorption (using
\texttt{xsphabs}; the parameters changed only minutely with other
choices, such as \texttt{xstbabs} and \texttt{xswabs}).  We tried NSA
models with and without a strong magnetic field; for both, we fixed
the mass and radius of the neutron star at $1.4\,M_\odot$ and 10\,km,
respectively.  Our results are summarized in Table~\ref{tab:specfit}
(there and below, temperatures and radii refer to those measured by a
distant observer).

\begin{figure}
\plotone{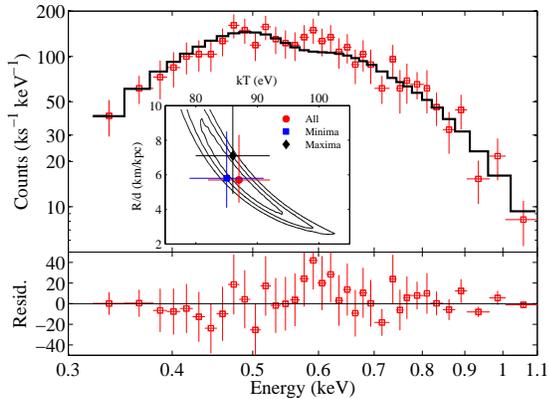}
\caption{The best-fit  blackbody model along with residuals
  (lower panel). Inset: best-fit contours of $kT$ vs.\ normalization $R/d$,
  with 1-, 2-, and 3-$\sigma$ joint confidence contours ($\Delta
  \chi^2=2.30,\,6.17\,11.8$).  The best-fit value is the red circle,
  while the blue square and black diamond are for the combined minima
  and maxima, respectively.}
\label{fig:spec}
\end{figure}

The NSA models gave reasonable fits, but implied implausibly small
distances: $33^{+22}_{-18}$ and $160^{+110}_{-60}$\,pc for the $B=0$
and $B=10^{13}\,$G models, respectively.  A power-law model gave a
significantly worse fit than the thermal models ($\chi^2=76.9$ vs.\
45.1 for 67 dof).  We found that, similar to the INSs, a simple
blackbody gave a good fit (Fig.~\ref{fig:spec}) and the most
reasonable parameters.\footnote{With the blackbody model and the
  observed 0.32--1.1\,keV count rate of $0.056{\rm\,s^{-1}}$, the
  expected count rate for ROSAT PSPC is $0.027{\rm\,s^{-1}}$
  (estimated using the Portable, Interactive Multi-Mission Simulator
  [PIMMS]; \url{http://heasarc.gsfc.nasa.gov/Tools/w3pimms.html}).
  Thus, most likely \PSR\ is responsible for all of the flux of \rxs.}
In what follows, we use that as our baseline model.  We find there is
significant covariance between the fitted parameters, which, within
$1\sigma$, leads to changes in absorption of $4\times10^{20}\,{\rm
  cm}^{-2}$ and in temperature of $\sim\!5$\,eV.  To explore the
influence of the absorption column, we also tried fits where we held
$N_{\rm H}$ fixed at 1 or 10 times the dispersion measure ($70\,{\rm
  cm}^{-3}\,{\rm pc}=\expnt{2.1}{20}\,{\rm cm^{-2}}$).  These fits
were somewhat worse than those with $N_{\rm H}$ free, but still had
reduced $\chi^2$ less than 1.

To constrain any additional non-thermal emission, we tried adding a
power law to the blackbody.  We found it had negligible effect on the
parameters, and, assuming $\Gamma=3$, we infer a $1\sigma$ upper limit
of the photon rate of $7\times10^{-6}\,{\rm s}^{-1}\,{\rm
  cm}^{-2}\,{\rm keV}^{-1}$
 at 1\,keV, corresponding to
$\lesssim\!3\%$ of the flux in the 0.32--1.1\,keV band.

We also tried fitting the spectrum of \PSR\ with a model that included
an additive Gaussian line, in order to constrain the presence of
absorption features similar to those seen for the INSs, which have
equivalent widths of up to several 100\,eV \citep{haberl07,vkk07}.  We
fixed the full width at half maximum (FWHM) of the line to 0.1\,keV,
comparable to the energy resolution of the back-illuminated ACIS-S3
CCD, and varied the line energy between 0.3\,keV and 1.1\,keV, fitting
only for its amplitude (allowing both positive and negative
amplitudes).  As expected from Fig.~\ref{fig:spec}, we did not find
evidence for any absorption line.  The largest improvement in the fit
occurs near 0.6\,keV, for an emission line with a height of $35$\%,
but even this reduces $\chi^2$ by only 4, which is not significant
given our number of trials (based on simulations, we estimate a
false-alarm probability of $\sim\!30\%$).  We infer a 68\%-confidence
upper limit of 50\,eV to the equivalent width of any feature.

Finally, we looked for phase-resolved trends in two ways.  First, we
determined the median energy as a function of pulse phase (upper panel
in Figure~\ref{fig:pulse}), finding little change except perhaps a
hint of hardening at the maximum that is coincident with the radio
peak, although given the current statistics this is not significant
(the $\chi^2$ for the median energy is 9 for 15 degrees-of-freedom).
Second, we separated the data into two bins: one containing the two
maxima and one the two minima, each with 50\% of the exposure time.
From our fit (Table~\ref{tab:specfit} and Figure~\ref{fig:spec}), we
find negligible difference in temperature between the maxima and
minima, consistent with the lack of change seen in median energy.
Thus, the pulsations appear to reflect mostly changes in emitting
area.

\subsection{Spatial Analysis}
To check for possible extended emission, e.g., from a pulsar wind
nebula, we compared our observations with a simulated point-spread
function (PSF), created using \texttt{ChaRT} (the \textit{Chandra} Ray
Tracer) and \texttt{MARX} (version 4.5.0), for a spectrum consistent
with our best-fit blackbody.  We find good agreement for radii
$\geq2\,$pixels ($1\arcsec$), although with only $\approx 180$ counts
at these radii, the constraints are not strong.  Oddly, the two bins
at the smallest radii, 0--0.5\,pixels and 0.5--1.0\,pixels, had some
deviations, with the model PSF underpredicting the counts in the first
bin by about 6\,$\sigma$ (30\%), and overpredicting the second by
about 4\,$\sigma$ (26\%).  Since it is unlikely that the pulsar is
more centrally concentrated than a point-source, we probably are
simply reaching the limits in the model (we experimented with the
\chandra\ processing, turning off pixel position randomization, but
found that this made little difference).  We conclude that there is no
evidence for extended emission on scales of $\gtrsim\!1\arcsec$.

\section{Discussion}
Regardless of the path by which \PSR\ and the INSs got to their present
states, we can examine how closely their current properties resemble
each other and whether the unique information from either helps
understand the thermal emission from this class of neutron stars as a
whole.  Furthermore, we can see how \PSR\ might help us understand the
evolution of high-$B$ pulsars and the origin of the INSs.  We consider
these in turn.

\subsection{Comparison with the INSs}

\PSR\ was selected to be similar to the INSs in having a long
spin-period, $P>3\,$s, and strong magnetic field,
$B=\expnt{(1-3)}{13}\,$G.  With a thermal spectrum with
$kT=87\pm5{\rm\,eV}$, and $R\approx6{\rm\,km}$ (at the nominal
distance of $\sim\!1\,$kpc in the \object{Gould Belt}), its X-ray
properties are similar as well (indeed, it would be classified as an
INS based on these).  Its spectrum has no evidence for absorption,
unlike \rbs, which has strong absorption \citep{hsh+03} while having a
very similar magnetic field and a slightly higher temperature
($B=\expnt{3.4}{13}$\,G, \citealt{kvk05b}; $kT=102$\,eV,
\citealt{shhm07}).  However, \rxjk, another INS with a similar if
slightly lower field ($B=\expnt{2.4}{13}$\,G, \citealt{kvk05}), had
weak absorption, with an equivalent width of $\sim\!5{\rm\,eV}$, which
we would not be able to detect, when its temperature was most similar
to \PSR\ ($kT=86{\rm\,eV}$, \citealt{hztb04}).  Intriguingly, this
source changed, developing much stronger absorption while becoming
hotter ($kT=94{\rm\,eV}$, \citealt{dvvmv04,htdv+06,vkkpm07}).  Taking
our inferred temperature at face value, it is thus possible that \PSR\
emits identically to an INS of the same temperature and dipole
magnetic field strength.  If so, at higher sensitivity it may show
weak absorption at $\sim\!300{\rm\,eV}$.  (We caution, however, that
the dependencies of absorption energy and strength on temperature and
magnetic field strength remain puzzling for the INSs as a class;
\citealt{kvk09,kvk09b}.)

\begin{deluxetable*}{ l c c c c c c}
\tablewidth{0pt}
\setlength{\tabcolsep}{2pt}
\tabletypesize{\footnotesize}
\tablecaption{Results of X-ray fits to PSR J0726$-$2612\label{tab:specfit}}
\tablehead{
\colhead{Model} & \colhead{$N_{\rm H}$} & \colhead{$\Gamma$ or $kT$} &
\colhead{$A$ or $R/d$\tablenotemark{a}} & \colhead{$F_{\rm abs}\tablenotemark{b}\times10^{-13}$} & \colhead{$F_{\rm
    unabs}\tablenotemark{b}\times10^{-13}$} & \colhead{$\chi^2/$DOF} \\
 & \colhead{($10^{20}\,{\rm cm}^{-2}$)} & \colhead{(N/A} &
\colhead{($10^{-5}\,{\rm s}^{-1}\,{\rm cm}^{-2}\,{\rm keV}^{-1}$} & 
\colhead{(${\rm erg\,s^{-1}\,cm^{-2}}$)} &
\colhead{(${\rm erg\,s^{-1}\,cm^{-2}}$)} \\
 & &\colhead{or eV)} & \colhead{or km\,kpc$^{-1}$)}
}
\startdata
Blackbody\dotfill & $\phn8_{-3}^{+4}$ & \phn\phantom{.}$87\pm5$\phn\phantom{.} & \phn$5.7_{-1.3}^{+2.6}$ & $4.5$ & $12.9$ & $45.1/67$ \\
Blackbody\dotfill & $\phn2_{\phantom{-}\phantom{0}}$ &
\phn\phantom{.}$95\pm2$\phn\phantom{.} &
\phn$3.5_{-0.2}^{+0.3}$ & 4.5 & \phn5.6 & 48.1/68\\
Blackbody\dotfill & $21_{\phantom{-}\phantom{0}}$ & \phn\phantom{.}$74\pm2$\phn\phantom{.} &
$17.1_{-1.4}^{+1.5}$ & 3.6 & 33.8 & 55.0/68\\
NSA $(B=0)$\dotfill & $14_{-4}^{+9}$ & \phn\phantom{.}$22\pm3$\phn\phantom{.} & $307_{-122}^{+370}$ & $4.2$ & $22.7$ & $52.8/67$ \\
NSA $(B=10^{13}\,{\rm G})$\dotfill & $12^{+4}_{-4}$ & \phn\phantom{.}$42\pm4$\phn\phantom{.} & \phantom{.}$63_{-24}^{+39\phantom{.}}$ & $4.3$ & $19.2$ & $52.8/67$  \\
Power Law ($\Gamma$ free)\dotfill & $20^{+4}_{-4}$ & \phn$6.4\pm0.4$ & \phn$7.6_{-0.8}^{+1.0}$ & $4.2$ & $60.6$ & $76.9/67$\\
Power Law ($\Gamma = 3$)\dotfill & $\phn\phantom{..}0^{+0.1}$ & $3$ & \phn$9.0^{+0.7}_{-0.2}$ & $3.2$ & \phn$3.2$ & $195/66$ \\
\multicolumn{2}{l}{Blackbody phase-resolved:}\\
Minima\tablenotemark{c}\dotfill & $\phn9_{-4}^{+5}$ &
\phn\phantom{.}$85\pm6$\phn\phantom{.}  & \phn$5.8_{-1.7}^{+2.7}$ & $3.3$ & \phn$8.5$ & $24.3/37$ \\
Maxima\tablenotemark{c}\dotfill & $\phn9_{-4}^{+5}$ & \phn\phantom{.}$86\pm6$\phn\phantom{.} & \phn$7.1_{-2.2}^{+3.5}$ & $5.1$ & $13.2$ & $47.1/51$ \\
\enddata
\tablecomments{Quantities without uncertainties were held fixed during
  the fit.  All uncertainties are for 1-$\sigma$ confidence, and all
  other parameters were allowed to vary during the calculation of the uncertainties.}
\tablenotetext{a}{For the NSA models, the mass is fixed to
  $1.4\,M_{\odot}$ and the radius is fixed at 10\,km, so the
  uncertainties calculated for  $R/d$  are based on the
  best-fit distance.  The normalization of the power-law model $A$ is
  the photon rate at an energy of 1\,keV.}
\tablenotetext{b}{The fluxes are given in the 0.32--1.1\,keV band and
  are given both as observed and corrected for absorption.}
\tablenotetext{c}{Phase-resolved spectroscopy using a blackbody
  model.  The two phase bins each include 50\% of the data, and were
fit simultaneously with the same absorption.}
\end{deluxetable*}

The inferred luminosity $L_{\rm X} = 1.5\times10^{32}d_{\rm
  kpc}^{2}{\rm\,erg\,s}^{-1}$ (0.32--1.1\,keV) is about half the
rotational energy loss rate $\dot{E}=4\pi^2I\dot{P}/P^3 =
\expnt{2.8}{32}\,{\rm erg\,s}^{-1}$ (for a moment of inertia
$I=10^{45}\,{\rm g\,cm}^2$).  This is on the low side of what is seen
for 
the INSs ($L_{\rm X}\sim (0.5\ldots 100)\dot E$), possibly just a consequence of
\PSR\ having a shorter period for the same $B$ (for dipole spin-down,
$\dot E\propto B^2/P^4$).  It is much larger than the non-thermal
emission typically observed for rotation-powered pulsars ($L_{\rm
  X}\approx10^{-3}\dot E$, \citealt{bt97}), although we cannot exclude
that \PSR\ has a similar non-thermal component: our 1$\sigma$ upper
limit of 3\% of the 0.32--1.1\,keV flux for the contribution of a
power-law with $\Gamma=3$ (a value typical for pulsars) implies
$L_{\rm PL}\lesssim 0.015d_{\rm kpc}^2\dot E$.

The pulse properties of \PSR\ also echo those of the INSs, which show
smooth profiles dominated by the fundamental or the first harmonic.
The pulse fraction of $27\pm5$\% is a little larger than the $\sim\!1$
to 18\% seen for the INSs \citep{haberl07}, but it is not dramatically
different.  We note that without the radio period, we might have
actually identified the pulse period of \PSR\ as half of the true
value.  The lack of temperature variation is somewhat strange (cf.\ up
to 10\% changes in $kT$ over the pulse of \rbs, for example;
\citealt{shhm05}), as, naively, it implies that we are seeing changes
in projected area of large regions of similar temperature with
anything in between so much colder that it does not contribute.  It
may be that instead the whole surface emits at more or less a uniform
temperature, and that the pulsations reflect anisotropies in the
emission, perhaps associated with the strong magnetic field.  The same
may hold for the INSs, which typically also show only modest changes
in temperature with pulse phase.

A possible difference with the INSs is that \PSR\ has an
order-of-magnitude smaller characteristic age.  This may again be a
consequence of its shorter period for the same $B$ (for dipole
spin-down, $\tau\propto(P/B)^2$), but might also point to an
evolutionary difference (see below).  Another characteristic of INSs is
that they have faint optical counterparts, but with optical/UV fluxes
a factor of 6--50 above the extrapolation of the X-rays.  This is in
contrast to middle-aged pulsars such as \object{PSR B0656+14} and
\object{Geminga}, which have ultraviolet fluxes that are more
consistent with the X-ray extrapolations \citep{ssl+05,kpzr05}.  It
will be interesting to see whether \PSR\ has an excess or not.

Of course, the most glaring difference between \PSR\ and the INSs is
the radio emission.  Numerous searches have yet to find confirmed
radio emission (coherent or bursty) from the INSs (\citealt{kml+09},
and references therein).  This may be a result of their location near
the pulsar ``death line,'' or more simply a consequence of their
narrow radio beams: the 50\% width of the radio pulse from \PSR\ is
$<1\%$ of the pulse period (consistent with general trends for
long-period pulsars) so the chance of missing the radio pulse is
large.

Overall, we conclude that with the exceptions of its radio emission
and characteristic age, the properties of \PSR\ agree with those of
the INSs.  Based on the compilations of \citet{kvk09b}, \citet{zkm+11},
and references therein, it is the best pulsar analog to the INSs to
date.

\subsection{Implications for the Evolution of High-$B$ Neutron Stars}

In order to look for evolutionary trends, we can compare \PSR\ to
sources with similar properties in the $P$-$\dot P$ diagram, as well
as with the evolution expected from the work of \cite{pmg09}, which
seems to explain the origins of the INSs (Fig.~\ref{fig:ppdot}).  A
first comparison is with INSs of similar field and temperature, \rxjk\
and \rbs.  As mentioned, the main difference is that the
characteristic age of \PSR\ is $200\,$kyr, while those of the INSs are
$>2\,$Myr.  This could imply that the true age of \PSR\ is closer to
its characteristic age than is the case for the INSs, and maybe that it
has undergone less $B$ decay.  This is consistent with the models of
\cite{pmg09}: $\expnt{3}{13}\,$G is near the dividing line where
field-decay becomes important, and the true age is expected to be no
more than a factor~2 shorter than the characteristic age (see
Fig.~\ref{fig:ppdot}).  Taking the models at face value, one infers
that \PSR\ was born with a dipole field of just below $10^{14}\,$G,
while most INSs would have had a field about a factor~2 stronger.

As a second comparison, we can compare \PSR\ with two X-ray-detected
objects that have almost the same period: the modest-$B$ INS \rxj\
($\expnt{1.0}{13}\,$G, \citealt{kvk11}), and the high-$B$ pulsar
\psrb\ ($\expnt{7.4}{13}\,$G, \citealt{zkm+11}).  The former is very
similar to \PSR, with the main difference being that it is cooler
($kT\approx 45\,$eV).  From the evolutionary tracks, it seems most
likely \rxj\ initially simply had even lower magnetic field strength,
which decayed by a factor $\sim\!2$ or so, and that, like \PSR, its
true age is at most a factor 2 shorter than its characteristic age of
2\,Myr.

From the evolutionary tracks, \psrb\ would be a good candidate
progenitor for INSs like \rxjk, and indeed it has a largely thermal
spectrum consistent with a blackbody with a significantly hotter
temperature ($kT\approx 190\,$eV), suggestive of on-going field decay.
However, the emitting radius of $\sim\!2\,$km and pulsed fraction of
$\sim\!50$\% differ from those found for the INSs (and \PSR), and are
more reminiscent of even younger high-$B$ pulsars like \object[PSR
J1119-6127]{PSR J1119$-$6127} \citep{gkc+08}, which also have high
pulsed fractions and smaller radii (although distance and fitting
uncertainties make the radii for all these objects poorly
constrained).  It may be that this reflects some combination of
ongoing $B$ decay and non-thermal heating; all these younger,
strong-field objects have substantially larger spin-down energy
losses.  Alternatively, the stronger field may lead to more
anisotropic emission.


Overall, it seems that the INSs and high-$B$ pulsars form a single
family (also see \citealt{kaspi10} for further discussion), with initial fields that were
stronger than their current ones, and that caused rapid initial
spin-down.  The long periods for the slower spinning objects also
imply low rotational energy losses and thus weak non-thermal emission.
Combined with some additional heating due to field decay, this makes
the thermal emission stand out more than it would in normal pulsars.

A continuing puzzle, however, is the difference with the magnetars
\citep{nk11}, which occupy similar parts of the $P$-$\dot P$ diagram,
but are clearly much more strongly affected by magnetic field decay.
It may be that those were born with much stronger toroidal components
to the field (which affect the spin-down only indirectly, via the more
rapid field decay).  In any case, there may be a smooth continuum,
given objects like \object[PSR J1846-0258]{PSR J1846$-$0258} ($B =
\expnt{4.9}{13}$\, G) which exhibited a sudden, magnetar-like X-ray
outburst \citep{ggg+08, ksh08, nsgh08}, the INS \rxjk\ that exhibited
a possible magnetic reconfiguration \citep{dvvmv04,vkkpm07}, the
transient magnetar \object[XTE J1810-197]{XTE~J1810$-$197} that was
detected in outburst \citep{ims+04} but in quiescence appears more
like an INS \citep{bpg+11}, or the putative low-$B$ magnetar
\object{SGR 0418+5729} ($<\expnt{8}{12}\,$G; \citealt{ret+10,tzp+11}).

\section{Outlook}

Our \chandra\ observations show that the relatively nearby high-$B$
pulsar \psr\ has properties similar to those of the INSs, showing
analogous thermal emission and similar smooth pulsations.  Because of
its radio emission, \psr\ may help us clarify the nature of the
emission and the pulsations.  First, for many radio pulsars the
observed polarization position angle curve is well described by the
rotating vector model \citep{rc69}, in which the position angle is assumed to be
aligned with a dipolar magnetic field.  Using this model, one can
infer the (mis)alignment between the spin and magnetic axes, as well
as the angle between the spin axis and line-of-sight.  Using such a
model for \PSR\ and combining it with fits to the X-ray light-curve
will help resolve the ambiguities that limited lightcurve analyses for
INSs \citep{br02,zt06,ho07}.  Second, unlike for the INSs where only
X-ray and optical astrometry (both of limited precision) are possible,
for \PSR\ traditional radio techniques are available (proper motions
with the Very Large Array, and potentially parallaxes with the Very
Long Baseline Array).  This means that even though the likely distance
is $\sim\!1\,$kpc, which would be well outside of reach for an optical
parallax, we can hope for a geometric distance measurement for \PSR\
that would constrain its emitting radius.



\acknowledgements We thank the referee for helpful comments.  Support
for this work was provided by the US National Aeronautics and Space
Administration (NASA) through Chandra award GO1-12080X and grant
NNX08AX39G, and by the Canadian Natural Sciences and Engineering
Research Council (NSERC).  This paper includes archived data obtained
through the Australia Telescope Online Archive and the CSIRO Data
Access Portal (http://datanet.csiro.au/dap/).  We thank W.~van~Straten
for assistance in interpreting those data.  We made extensive use of
SIMBAD and ADS.

{\it Facilities:} \facility{CXO (ACIS)}


\begin{thebibliography}{}

\bibitem[{Becker} \& {Tr\"{u}mper}(1997){Becker} \& {Tr\"{u}mper}]{bt97}
{Becker}, W. \& {Tr\"{u}mper}, J. 1997, \aap, 326, 682

\bibitem[{Bernardini} {et~al.}(2011){Bernardini}, {Perna}, {Gotthelf},  {Israel}, {Rea}, \& {Stella}]{bpg+11}
{Bernardini}, F., {Perna}, R., {Gotthelf}, E., {Israel}, G.~L., {Rea}, N., \&  {Stella}, L. 2011, ArXiv e-prints

\bibitem[{Braje} \& {Romani}(2002){Braje} \& {Romani}]{br02}
{Braje}, T.~M. \& {Romani}, R.~W. 2002, \apj, 580, 1043

\bibitem[{Buccheri} {et~al.}(1983){Buccheri}, {Bennett}, {Bignami}, {Bloemen},  {Boriakoff}, {Caraveo}, {Hermsen}, {Kanbach}, {Manchester}, {Masnou},  {Mayer-Hasselwander}, {Ozel}, {Paul}, {Sacco}, {Scarsi}, \&  {Strong}]{bbb+83}
{Buccheri}, R., {et al.} 1983,  \aap, 128, 245

\bibitem[{Burgay} {et~al.}(2006){Burgay} {et~al.}]{burg+06}
{Burgay} {et~al.} 2006, \mnras, 368, 283

\bibitem[Cordes \& Lazio(2002)Cordes \& Lazio]{cl02}
Cordes \& Lazio. 2002, astro-ph/0207156

\bibitem[{de Vries} {et~al.}(2004){de Vries}, {Vink}, {M{\' e}ndez}, \&  {Verbunt}]{dvvmv04}
{de Vries}, C.~P., {Vink}, J., {M{\' e}ndez}, M., \& {Verbunt}, F. 2004, \aap,  415, L31

\bibitem[{Faucher-Gigu{\`e}re} \& {Kaspi}(2006){Faucher-Gigu{\`e}re} \& {Kaspi}]{fgk06}
{Faucher-Gigu{\`e}re}, C.-A. \& {Kaspi}, V.~M. 2006, \apj, 643, 332

\bibitem[{Garmire} {et~al.}(2003){Garmire}, {Bautz}, {Ford}, {Nousek}, \&  {Ricker}]{gbf+03}
{Garmire}, G.~P., {Bautz}, M.~W., {Ford}, P.~G., {Nousek}, J.~A., \& {Ricker},  G.~R. 2003, \procspie, 4851, 28

\bibitem[{Gavriil} {et~al.}(2008){Gavriil}, {Gonzalez}, {Gotthelf}, {Kaspi},  {Livingstone}, \& {Woods}]{ggg+08}
{Gavriil}, F.~P., {Gonzalez}, M.~E., {Gotthelf}, E.~V., {Kaspi}, V.~M.,  {Livingstone}, M.~A., \& {Woods}, P.~M. 2008, Science, 319, 1802

\bibitem[{Gonzalez} {et~al.}(2005){Gonzalez}, {Kaspi}, {Camilo}, {Gaensler},  \& {Pivovaroff}]{gkc+08}
{Gonzalez}, M.~E., {Kaspi}, V.~M., {Camilo}, F., {Gaensler}, B.~M., \&  {Pivovaroff}, M.~J. 2005, \apj, 630, 489

\bibitem[{Gonzalez} {et~al.}(2004){Gonzalez}, {Kaspi}, {Lyne}, \&  {Pivovaroff}]{gkl+04}
{Gonzalez}, M.~E., {Kaspi}, V.~M., {Lyne}, A.~G., \& {Pivovaroff}, M.~J. 2004,  \apjl, 610, L37

\bibitem[{Haberl}(2007){Haberl}]{haberl07}
{Haberl}, F. 2007, \apss, 308, 181

\bibitem[{Haberl} {et~al.}(2003){Haberl}, {Schwope}, {Hambaryan}, {Hasinger},  \& {Motch}]{hsh+03}
{Haberl}, F., {Schwope}, A.~D., {Hambaryan}, V., {Hasinger}, G., \& {Motch}, C.  2003, \aap, 403, L19

\bibitem[{Haberl} {et~al.}(2006){Haberl}, {Turolla}, {de Vries}, {Zane},  {Vink}, {M{\'e}ndez}, \& {Verbunt}]{htdv+06}
{Haberl}, F., {Turolla}, R., {de Vries}, C.~P., {Zane}, S., {Vink}, J.,  {M{\'e}ndez}, M., \& {Verbunt}, F. 2006, \aap, 451, L17

\bibitem[{Haberl} {et~al.}(2004){Haberl}, {Zavlin}, {Tr{\" u}mper}, \&  {Burwitz}]{hztb04}
{Haberl}, F., {Zavlin}, V.~E., {Tr{\" u}mper}, J., \& {Burwitz}, V. 2004, \aap,  419, 1077

\bibitem[{Ho}(2007){Ho}]{ho07}
{Ho}, W.~C.~G. 2007, \mnras, 380, 71

\bibitem[{Ibrahim} {et~al.}(2004){Ibrahim}, {Markwardt}, {Swank}, {Ransom},  {Roberts}, {Kaspi}, {Woods}, {Safi-Harb}, {Balman}, {Parke}, {Kouveliotou},  {Hurley}, \& {Cline}]{ims+04}
{Ibrahim}, A.~I., {et al.} 2004, \apjl, 609, L21

\bibitem[{Kaplan}(2008){Kaplan}]{kaplan08}
{Kaplan}, D.~L. 2008, AIPC, 983, 331, arXiv:0801.1143

\bibitem[{Kaplan} \& {van Kerkwijk}(2005a){Kaplan} \& {van Kerkwijk}]{kvk05}
{Kaplan}, D.~L. \& {van Kerkwijk}, M.~H. 2005a, \apjl, 628, L45

\bibitem[{Kaplan} \& {van Kerkwijk}(2005b){Kaplan} \& {van Kerkwijk}]{kvk05b}
---. 2005b, \apjl, 635, L65

\bibitem[{Kaplan} \& {van Kerkwijk}(2009a){Kaplan} \& {van Kerkwijk}]{kvk09}
---. 2009a, \apjl, 692, L62

\bibitem[{Kaplan} \& {van Kerkwijk}(2009b){Kaplan} \& {van Kerkwijk}]{kvk09b}
---. 2009b, \apj, 705, 798

\bibitem[{Kaplan} \& {van Kerkwijk}(2011){Kaplan} \& {van Kerkwijk}]{kvk11}
---. 2011, \apjl, 740, L30

\bibitem[{Kaplan} {et~al.}(2007){Kaplan}, {van Kerkwijk}, \&  {Anderson}]{kvka07}
{Kaplan}, D.~L., {van Kerkwijk}, M.~H., \& {Anderson}, J. 2007, \apj, 660, 1428

\bibitem[{Kargaltsev} {et~al.}(2005){Kargaltsev}, {Pavlov}, {Zavlin}, \&  {Romani}]{kpzr05}
{Kargaltsev}, O.~Y., {Pavlov}, G.~G., {Zavlin}, V.~E., \& {Romani}, R.~W. 2005,  \apj, 625, 307

\bibitem[{Kaspi} \& {McLaughlin}(2005){Kaspi} \& {McLaughlin}]{km05}
{Kaspi} \& {McLaughlin}. 2005, \apjl, 618, L41

\bibitem[{Kaspi}(2010){Kaspi}]{kaspi10}
{Kaspi}, V.~M. 2010, Proceedings of the National Academy of Science, 107, 7147

\bibitem[{Kondratiev} {et~al.}(2009){Kondratiev}, {McLaughlin}, {Lorimer},  {Burgay}, {Possenti}, {Turolla}, {Popov}, \& {Zane}]{kml+09}
{Kondratiev}, V.~I., {McLaughlin}, M.~A., {Lorimer}, D.~R., {Burgay}, M.,  {Possenti}, A., {Turolla}, R., {Popov}, S.~B., \& {Zane}, S. 2009, \apj, 702,  692

\bibitem[{Kumar} \& {Safi-Harb}(2008){Kumar} \& {Safi-Harb}]{ksh08}
{Kumar}, H.~S. \& {Safi-Harb}, S. 2008, \apjl, 678, L43

\bibitem[{Manchester} {et~al.}(2005){Manchester}, {Hobbs}, {Teoh}, \&  {Hobbs}]{mhth05}
{Manchester}, R.~N., {Hobbs}, G.~B., {Teoh}, A., \& {Hobbs}, M. 2005, \aj, 129,  1993

\bibitem[{McLaughlin} {et~al.}(2007){McLaughlin}, {Rea}, {Gaensler},  {Chatterjee}, {Camilo}, {Kramer}, {Lorimer}, {Lyne}, {Israel}, \&  {Possenti}]{mrg+07}
{McLaughlin}, M.~A., {et al.} 2007, \apj, 670, 1307

\bibitem[{Motch} {et~al.}(2009){Motch}, {Pires}, {Haberl}, {Schwope}, \&  {Zavlin}]{mph+09}
{Motch}, C., {Pires}, A.~M., {Haberl}, F., {Schwope}, A., \& {Zavlin}, V.~E.  2009, \aap, 497, 423

\bibitem[{Motch} {et~al.}(2005){Motch}, {Sekiguchi}, {Haberl}, {Zavlin},  {Schwope}, \& {Pakull}]{msh+05}
{Motch}, C., {Sekiguchi}, K., {Haberl}, F., {Zavlin}, V.~E., {Schwope}, A., \&  {Pakull}, M.~W. 2005, \aap, 429, 257

\bibitem[{Ng} \& {Kaspi}(2011){Ng} \& {Kaspi}]{nk11}
{Ng}, C.-Y. \& {Kaspi}, V.~M. 2011, in American Institute of Physics Conference  Series, Vol. 1379, American Institute of Physics Conference Series, ed.  {E.~G{\"o}{\u g}{\"u}{\c s}, T.~Belloni, {\ Uuml}.~Ertan }, 60--69

\bibitem[{Ng} {et~al.}(2008){Ng}, {Slane}, {Gaensler}, \& {Hughes}]{nsgh08}
{Ng}, C.-Y., {Slane}, P.~O., {Gaensler}, B.~M., \& {Hughes}, J.~P. 2008, \apj,  686, 508

\bibitem[{Pavlov} {et~al.}(1991){Pavlov}, {Shibanov}, \& {Zavlin}]{psz91}
{Pavlov}, G.~G., {Shibanov}, I.~A., \& {Zavlin}, V.~E. 1991, \mnras, 253, 193

\bibitem[{Pons} {et~al.}(2009){Pons}, {Miralles}, \& {Geppert}]{pmg09}
{Pons}, J.~A., {Miralles}, J.~A., \& {Geppert}, U. 2009, \aap, 496, 207

\bibitem[{Popov} {et~al.}(2010){Popov}, {Pons}, {Miralles}, {Boldin}, \&  {Posselt}]{ppm+10}
{Popov}, S.~B., {Pons}, J.~A., {Miralles}, J.~A., {Boldin}, P.~A., \&  {Posselt}, B. 2010, \mnras, 401, 2675

\bibitem[{Popov} {et~al.}(2005){Popov}, {Turolla}, {Prokhorov}, {Colpi}, \&  {Treves}]{ptp+05}
{Popov}, S.~B., {Turolla}, R., {Prokhorov}, M.~E., {Colpi}, M., \& {Treves}, A.  2005, \apss, 299, 117

\bibitem[{Radhakrishnan} \& {Cooke}(1969){Radhakrishnan} \& {Cooke}]{rc69}
{Radhakrishnan}, V. \& {Cooke}, D.~J. 1969, \aplett, 3, 225

\bibitem[{Ransom}(2001){Ransom}]{ransom01}
{Ransom}, S.~M. 2001, PhD thesis, Harvard University

\bibitem[{Rea} {et~al.}(2010){Rea}, {Esposito}, {Turolla}, {Israel}, {Zane},  {Stella}, {Mereghetti}, {Tiengo}, {G{\"o}tz}, {G{\"o}{\u g}{\"u}{\c s}}, \&  {Kouveliotou}]{ret+10}
{Rea}, N., {et al.} 2010, Science, 330, 944

\bibitem[Refsdal {et~al.}(2009)Refsdal, Doe, Nguyen, Siemiginowska,  Bonaventura, Burke, Evans, Evans, Fruscione, Galle, Houck, Karovska, Lee, \&  Nowak]{sherpa}
Refsdal, B., {et al.} 2009, in Proceedings of the 8th Python in Science  conference (SciPy 2009), ed. G.~Varoquaux, S.~van~der Walt, \& J.~Millman,  51--57

\bibitem[{Schwope} {et~al.}(2005){Schwope}, {Hambaryan}, {Haberl}, \&  {Motch}]{shhm05}
{Schwope}, A.~D., {Hambaryan}, V., {Haberl}, F., \& {Motch}, C. 2005, \aap,  441, 597

\bibitem[{Schwope} {et~al.}(2007){Schwope}, {Hambaryan}, {Haberl}, \&  {Motch}]{shhm07}
---. 2007, \apss, 308, 619

\bibitem[{Shibanov} {et~al.}(2005){Shibanov}, {Sollerman}, {Lundqvist},  {Gull}, \& {Lindler}]{ssl+05}
{Shibanov}, Y.~A., {Sollerman}, J., {Lundqvist}, P., {Gull}, T., \& {Lindler},  D. 2005, \aap, 440, 693

\bibitem[{Tetzlaff} {et~al.}(2011){Tetzlaff}, {Eisenbeiss}, {Neuhaeuser}, \&  {Hohle}]{tenh11}
{Tetzlaff}, N., {Eisenbeiss}, T., {Neuhaeuser}, R., \& {Hohle}, M.~M. 2011,  \mnras, in press, arXiv:1107.1673

\bibitem[{Tetzlaff} {et~al.}(2010){Tetzlaff}, {Neuh{\"a}user}, {Hohle}, \&  {Maciejewski}]{tnhm10}
{Tetzlaff}, N., {Neuh{\"a}user}, R., {Hohle}, M.~M., \& {Maciejewski}, G. 2010,  \mnras, 402, 2369

\bibitem[{Turolla} {et~al.}(2011){Turolla}, {Zane}, {Pons}, {Esposito}, \&  {Rea}]{tzp+11}
{Turolla}, R., {Zane}, S., {Pons}, J.~A., {Esposito}, P., \& {Rea}, N. 2011,  \apj, in press, arXiv:1107.5488

\bibitem[{van Kerkwijk} \& {Kaplan}(2007){van Kerkwijk} \& {Kaplan}]{vkk07}
{van Kerkwijk}, M.~H. \& {Kaplan}, D.~L. 2007, \apss, 308, 191

\bibitem[{van Kerkwijk} \& {Kaplan}(2008){van Kerkwijk} \& {Kaplan}]{vkk08}
---. 2008, \apjl, 673, L163

\bibitem[{van Kerkwijk} {et~al.}(2007){van Kerkwijk}, {Kaplan}, {Pavlov}, \&  {Mori}]{vkkpm07}
{van Kerkwijk}, M.~H., {Kaplan}, D.~L., {Pavlov}, G.~G., \& {Mori}, K. 2007,  \apjl, 659, L149

\bibitem[{Voges} {et~al.}(1999){Voges} {et~al.}]{rbs}
{Voges}, W. {et~al.} 1999, \aap, 349, 389

\bibitem[{Walter}(2001){Walter}]{wal01}
{Walter}, F.~M. 2001, \apj, 549, 433

\bibitem[{Weisskopf} {et~al.}(2000){Weisskopf}, {Tananbaum}, {Van Speybroeck},  \& {O'Dell}]{wtvso00}
{Weisskopf}, M.~C., {Tananbaum}, H.~D., {Van Speybroeck}, L.~P., \& {O'Dell},  S.~L. 2000, \procspie, 4012, 2

\bibitem[{Zane} \& {Turolla}(2006){Zane} \& {Turolla}]{zt06}
{Zane}, S. \& {Turolla}, R. 2006, \mnras, 366, 727

\bibitem[{Zavlin} {et~al.}(1996){Zavlin}, {Pavlov}, \& {Shibanov}]{zps96}
{Zavlin}, V.~E., {Pavlov}, G.~G., \& {Shibanov}, Y.~A. 1996, \aap, 315, 141

\bibitem[{Zhu} {et~al.}(2008){Zhu}, {Kaspi}, {Dib}, {Woods}, {Gavriil}, \&  {Archibald}]{zkd+08}
{Zhu}, W., {Kaspi}, V.~M., {Dib}, R., {Woods}, P.~M., {Gavriil}, F.~P., \&  {Archibald}, A.~M. 2008, \apj, 686, 520

\bibitem[{Zhu} {et~al.}(2011){Zhu}, {Kaspi}, {McLaughlin}, {Pavlov}, {Ng},  {Manchester}, {Gaensler}, \& {Woods}]{zkm+11}
{Zhu}, W.~W., {Kaspi}, V.~M., {McLaughlin}, M.~A., {Pavlov}, G.~G., {Ng},  C.-Y., {Manchester}, R.~N., {Gaensler}, B.~M., \& {Woods}, P.~M. 2011, \apj,  734, 44

\end{thebibliography}


\end{document}